\documentclass[aps,preprint,showpacs, showkeys]{revtex4}%
\usepackage{amsfonts}
\usepackage{amsmath}
\usepackage{amssymb}
\usepackage{graphicx}
\usepackage{float}%
\providecommand{\U}[1]{\protect\rule{.1in}{.1in}}

\begin{document}
\title[ ]{Effective Field calculations of the Energy Spectrum of the $\mathcal{PT}%
$-Symmetric ($-x^{4}$) Potential}
\author{${}^{1,2}$ Abouzeid M. Shalaby }
\email{amshalab@mans.edu.eg}
\affiliation{${}^{1}$Physics Department, Faculty of Science, Mansoura University, Egypt\\
${}^{2}$ Physics Department, Faculty of Science, Qassim University, KSA }
\keywords{non-Hermitian models, $\mathcal{PT}$-symmetric theories, effective potential.}
\pacs{03.65.-w, 11.10.Kk, 02.30.Mv, 11.30.Qc, 11.15.Tk}

\begin{abstract}
In this work, we show that the traditional effective field approach can be
applied to the $\mathcal{PT}$-symmetric wrong sign ($-x^{4}$) quartic
potential. The importance of this work lies in the possibility of its
extension to the more important $\mathcal{PT}$-symmetric quantum field theory
while the other approaches which use complex contours are not willing to be
applicable. We calculated the effective potential of the massless $-x^{4}$
theory as well as the full spectrum of the theory. Although the calculations
are carried out up to first order in the coupling, the predicted spectrum is
very close to the exact one taken from other works. The most important result
of this work is that the effective potential obtained, which is equivalent to
the Gaussian effective potential, is bounded from below while the classical
potential is bounded from above. This explains the stability of the vacuum of
the theory. The obtained quasi-particle Hamiltonian is non-Hermitian but
$\mathcal{PT}$-symmetric and we showed that the calculation of the metric
operator can go perturbatively. In fact, the calculation of the metric
operator can be done even for higher dimensions (quantum field theory) which,
up till now, can not be calculated in the other approaches either
perturbatively or in a closed form due to the possible appearance of field
radicals. Moreover, we argued that the effective theory is perturbative for
the whole range of the coupling constant and the perturbation series is
expected to converge rapidly (the effective coupling $g_{eff}=\frac{1}{6}$).

\end{abstract}
\maketitle

The pioneering article of Carl Bender and Stefan Boettcher \cite{bendr} puts
the pseudo Hermitian theories with real spectra on a track in which both
theories, Hermitian and pseudo Hermitian, are in the same footing in regard to
the physical acceptability. For pseudo Hermitian theories, the calculation of
a positive definite metric operator, or the $C$ operator, is indispensable. An
exception is noticed for theories in which the $Q$ operator represents a gauge
transformation \cite{qop}. For such theories, the $Q$ operator disappears from
the physical calculations. However, even for quantum mechanical theories in
which the metric operator is not a gauge transformation, there exists known
successful algorithms for the calculations of the metric or the $C$ operator.
On the other hand, for bounded from above scalar field theories like the
$-\phi^{4}$ theory, it seems that the conventional algorithms used in the
quantum mechanical case are inapplicable. For instance, the metric operator
has not been obtained yet either perturbatively or in a closed form. Besides,
Physical amplitudes have not been calculated in any regime for the $-\phi^{4}$
scalar field theory. Accordingly, we need a successful as well as aplicable
algorithm for the investigation of this theory which bears the interesting
asymptotic-freedom property.

\begin{figure}[th]
\begin{center}
\includegraphics{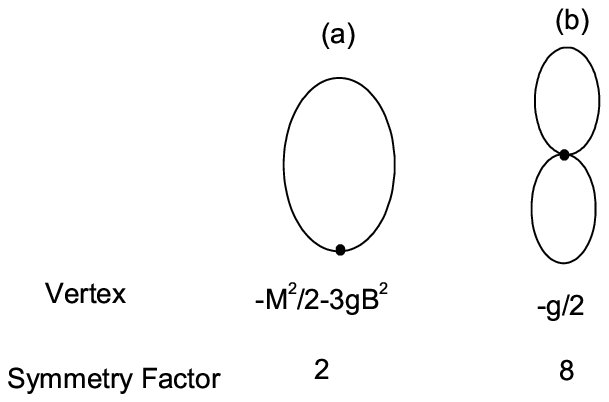}
\end{center}
\caption{The Feynman diagrams contributing to the first order effective
potential.}%
\label{feyn}%
\end{figure}

\begin{figure}[ptb]
\begin{center}
\includegraphics [width=8cm,height=8cm]{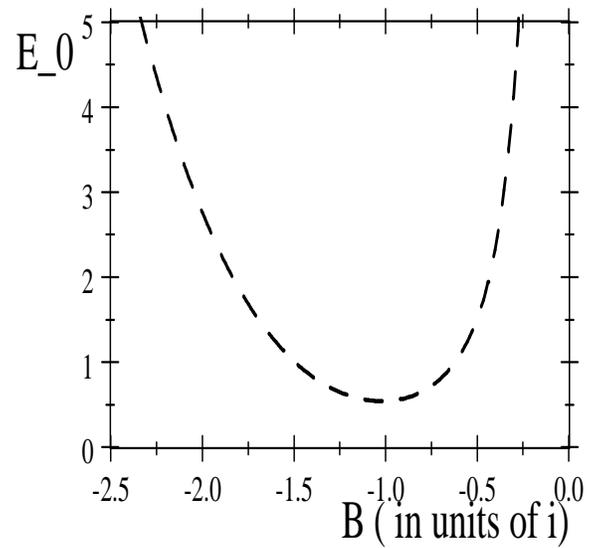}
\end{center}
\caption{The ground state energy as a function of the one pint function $B$
measured in units of $i$ for $g=0.5$ for the $\mathcal{PT}$-symmetric $-x^{4}$
potential.}%
\label{eob}%
\end{figure}

In the regime of $\mathcal{PT}$-symmetric theories, the quantum mechanical,
bounded from above, potentials are always treated in a complex contour which
seems to be inapplicable for quantum field cases because of the probable
appearance of a root of the field. Moreover, the theory is non-perturbative
and the perturbative calculations for the $Q$ operator, for instance, leaded
to a trivial $C$ operator. However, the traditional effective field technique,
as a non-perturbative tool, has not been advocated seriously in the regime of
$\mathcal{PT}$-symmetric theories. In fact, in previous works, we have used
effective field calculations to obtain the one particle irreducible ($1PI$)
amplitudes for the $-\phi^{4}$ theory in $1+1$ and $2+1$ dimensions \cite{AIB,
czv,PRD}. The results of Ref.\cite{czv} has reproduced the exponential
vanishing of the vacuum condensate as the coupling constant goes to zero as
predicted by a different technique used in Ref. \cite{bendcal} which assures
the success of the algorithm in non-Hermitian quantum field theories. What
makes this trend very impressive is that we keep an eye on the fruitful
features of the $\mathcal{PT-}$symmetric field theory to play the role of the
Higgs mechanism in the standard model. However, in this work, we aim to test
the validity of the algorithm in a quantitative manner and since there exist
rigorous numerical results for the quantum mechanical $-x^{4}$ model, we will
test the validity of the effective field approach for the massless quantum
mechanical Hamiltonian of the from;
\begin{equation}
H=\frac{1}{2}p^{2}-\frac{g}{2}x^{4}, \label{x4}%
\end{equation}
or in a quantum field language one may use the notations;
\[
H(x)=\frac{1}{2}\left(  \left(  \nabla\phi\right)  ^{2}+\pi^{2}\right)
-\frac{g}{2}\phi^{4},
\]
for the Hamiltonian density $H(x)$ with the calculation proceeds in $0+1$
dimensions (quantum mechanics). The effective field studies start by applying
the canonical transformation of the form;
\[
\phi=\psi+B,\text{ \ }\pi=\Pi=\dot{\psi},\text{\ \ \ }%
\]
where $B$ is the vacuum condensate and the fields $\psi$ and $\Pi$ follow from
the relations;
\begin{align}
\psi(x)  &  =%
{\displaystyle\int}
\frac{d\overrightarrow{k}}{\left(  2\pi\right)  ^{d}\sqrt{2w}}\left\{
a(\overrightarrow{k})\exp\left(  i\overrightarrow{k}.\overrightarrow
{x}-iwx_{o}\right)  +a^{\dagger}(\overrightarrow{k})\exp\left(
-i\overrightarrow{k}.\overrightarrow{x}+iwx_{o}\right)  \right\}  ,\nonumber\\
\Pi(x)  &  =\frac{1}{i}%
{\displaystyle\int}
\frac{d\overrightarrow{k}}{\left(  2\pi\right)  ^{d}}\sqrt{\frac{w}{2}%
}\left\{  a(\overrightarrow{k})\exp\left(  i\overrightarrow{k}.\overrightarrow
{x}-iwx_{o}\right)  -a^{\dagger}(\overrightarrow{k})\exp\left(
-i\overrightarrow{k}.\overrightarrow{x}+iwx_{o}\right)  \right\}  ,\nonumber\\
w(k)  &  =\sqrt{\overrightarrow{k}^{2}+M^{2}\text{ \ \ }}\text{ ,
\ \ \ \ \ \ \ \ \ }\left[  a(\overrightarrow{k}),a^{\dagger}(\overrightarrow
{k}^{\prime})\right]  =\delta(\overrightarrow{k},\overrightarrow{k}^{\prime}).
\label{fiweak}%
\end{align}
Also, $M$ is the mass of the field $\psi$ ($\phi$ is massless) and $d$ is the
dimension of the position space. Accordingly, the Hamiltonian density
transforms as
\begin{align}
H(x)  &  =\frac{1}{2}\left(  \left(  \nabla\psi\right)  ^{2}+\Pi^{2}\right)
-\frac{g}{2}\left(  \psi+B\right)  ^{4},\nonumber\\
&  =\frac{1}{2}\left(  \left(  \nabla\psi\right)  ^{2}+\Pi^{2}+M^{2}\psi
^{2}\right) \label{quasip}\\
&  -\frac{g}{2}\left(  \psi^{4}+4B\psi^{3}+\left(  6B^{2}\right)  \psi
^{2}+\left(  4B^{3}\right)  \psi+B^{4}\right)  -\frac{1}{2}M^{2}\psi
^{2},\nonumber\\
&  =\frac{1}{2}\left(  \left(  \nabla\psi\right)  ^{2}+\Pi^{2}+M^{2}\psi
^{2}\right)  -\frac{g}{2}\left(  \psi^{4}+4B\psi^{3}\right) \nonumber\\
&  +\left(  -\frac{1}{2}M^{2}-3gB^{2}\right)  \psi^{2}-2gB^{3}\psi-\frac{g}%
{2}B^{4}\\
&  =H_{0}+H_{I}-\frac{g}{2}B^{4},\nonumber
\end{align}
where \
\begin{align}
H_{0}  &  =\frac{1}{2}\left(  \left(  \nabla\psi\right)  ^{2}+\Pi^{2}%
+M^{2}\psi^{2}\right)  ,\nonumber\\
\ H_{I}  &  =-\frac{g}{2}\left(  \psi^{4}+4B\psi^{3}\right)  +\left(
-\frac{1}{2}M^{2}-3gB^{2}\right)  \psi^{2}-2gB^{3}\psi.
\end{align}
The effective potential is defined as the vacuum energy \cite{Peskin}. Before
we go into the calculation of the effective potential, we need to explain the
nature of the algorithm we use and its relation to other algorithms. Regarding
this, it is well known that an amplitude calculated with respect to the true
vacuum is converted into a calculation with respect to the free vacuum via the
insertion of time evolution operator \cite{Peskin}. This realization is very
interesting since with respect to the free field, the operator $-\frac{g}%
{2}\psi^{4}$ in the effective Hamiltonian in Eq.(\ref{quasip}) is Hermitian
and thus the theory is a real line one. Moreover, through the expansion of the
time evolution operator, one obtains the Feynman diagrams of different orders
in the perturbation series. With this in mind, the effective potential is
generated as the expectation value of the potential term $H_{I}-\frac{g}%
{2}B^{4}$ in Eq.(\ref{quasip}) plus the expectation value of the kinetic term.
Through the expansion of the time evolution operator one obtains
$U(t,t_{0})=1+(-i)\int_{t_{0}}^{t}dtH_{I}\left(  t_{1}\right)  +......$, where
$U(t,t_{0})$ is the time evolution operator. Accordingly, keeping only the
first term (first order of $V_{eff}$) we get $V_{eff}=\langle0|H|0\rangle$,
where $|0\rangle$ is the vacuum corresponding to the free field $\psi$.
\ Accordingly, the vacuum energy, up to first order, is given by;%
\begin{equation}
E_{0}=\langle0|H|0\rangle=\langle0|H_{0}+H_{I}-\frac{g}{2}B^{4}|0\rangle,
\end{equation}
which has the same form as the Gaussian effective potential (GEP) studied in
Ref.\cite{GEP}. However, the parameters $B$ and $M$ are fixed using the fact
that the effective potential is the generating functional of the one particle
irreducible amplitudes and thus we have the relations;%
\begin{equation}
\frac{\partial V_{eff}}{\partial B}=\frac{\partial E_{0}}{\partial B}=0\text{,
}\frac{\partial^{2}V_{eff}}{\partial B^{2}}=\frac{\partial^{2}E_{0}}{\partial
B^{2}}=M^{2}. \label{paramf}%
\end{equation}
Now, up to first order in the coupling $g$, one have the Feynman diagrams
shown in Fig. \ref{feyn}. The two diagrams will contribute ($D=d+0=1$);
\begin{align*}
\text{diagram (a) }  &  \equiv\left(
\begin{array}
[c]{c}%
\left(  -i\right)  \frac{2i}{2}\left(  -\frac{M^{2}}{2}-3gB^{2}\right)
{\displaystyle\int}
\frac{d^{D}p}{\left(  2\pi\right)  ^{D}}\frac{1}{p^{2}-M^{2}}\\
=\left(  -i\right)  ^{2}i\left(  -\frac{M^{2}}{2}-3gB^{2}\right)  \frac
{1}{\left(  4\pi\right)  ^{\frac{1}{2}}}\frac{\Gamma\left(  1-\frac{1}%
{2}\right)  }{\Gamma\left(  1\right)  }\left(  \frac{1}{-\left\vert
M\right\vert ^{2}}\right)  ^{1-\frac{1}{2}},\\
=-\frac{1}{2M}\left(  3gB^{2}+\frac{M^{2}}{2}\right)  ,
\end{array}
\right) \\
\text{diagram (b) }  &  \text{=}\left(  -i\right)  \frac{\left(  12ig\right)
}{8}\left(
{\displaystyle\int}
\frac{d^{D}p}{\left(  2\pi\right)  ^{D}}\frac{1}{p^{2}-M^{2}}\right)  ^{2}=\\
&  \left(  -i\right)  \frac{\left(  -12ig\right)  }{8}\left(  \frac{1}{\left(
4\pi\right)  ^{\frac{1}{2}}}\frac{\Gamma\left(  1-\frac{1}{2}\right)  }%
{\Gamma\left(  1\right)  }\left(  \frac{1}{-\left\vert M\right\vert ^{2}%
}\right)  ^{1-\frac{1}{2}}\right)  ^{2},\\
&  =-\frac{3}{8}\frac{g}{\left\vert M\right\vert ^{2}}.
\end{align*}
to the effective potential. Thus the ground state energy can be obtained as
\begin{equation}
E_{0}=\frac{1}{2}M-\frac{1}{2}g\frac{3}{4M^{2}}+\left(  -\frac{1}{2}%
M^{2}-3gB^{2}\right)  \frac{1}{2M}+\left(  -\frac{1}{2}B^{4}g\right)  .
\end{equation}
The regime of the effective potential relates the derivatives of the effective
potential \ with respect to the vacuum condensate $B$ to the one particle
irreducible amplitudes (1PI) \cite{Peskin}. In other words, up to first order
in the coupling, the effective potential is constrained by the following two
conditions;%
\begin{align}
\frac{\partial V_{eff}}{\partial B}  &  =\frac{\partial E_{0}}{\partial
B}=\frac{\partial}{\partial B}\left(
\begin{array}
[c]{c}%
\frac{1}{2}M-\frac{1}{2}g\left(  \frac{3}{4M^{2}}\right) \\
+\left(  -\frac{1}{2}M^{2}-3gB^{2}\right)  \frac{1}{2M}+\left(  -\frac{1}%
{2}B^{4}g\right)
\end{array}
\right)  =0,\nonumber\\
\frac{\partial^{2}V_{eff}}{\partial B^{2}}  &  =\frac{\partial^{2}E_{0}%
}{\partial B^{2}}=\frac{\partial^{2}}{\partial B^{2}}\left(
\begin{array}
[c]{c}%
\frac{1}{2}M-\frac{1}{2}g\left(  \frac{3}{4M^{2}}\right) \\
+\left(  -\frac{1}{2}M^{2}-3gB^{2}\right)  \frac{1}{2M}+\left(  -\frac{1}%
{2}B^{4}g\right)
\end{array}
\right)  =M^{2}, \label{bbl}%
\end{align}
or equivalently
\begin{align}
\left(  -2g\right)  B^{3}+\left(  -\frac{3}{M}g\right)  B  &  =0,\nonumber\\
\allowbreak\left(  -6g\right)  B^{2}-\frac{3}{M}g\allowbreak &  =M^{2}.
\end{align}
Note that, $E_{0}$ obtained here is equivalent to the GEP with the condition
$\frac{\partial^{2}E_{0}}{\partial B^{2}}$ is equivalent to the condition of
minimal sensitivity $\frac{\partial E_{0}}{\partial M}$ used in Ref.
\cite{GEP}. However, the conditions used here, $\frac{\partial E_{0}}{\partial
B}$ and $\frac{\partial^{2}E_{0}}{\partial B^{2}}$, are more illuminating as
they are constraining the predicated $B$ and $M$ values to represent a minimum
of the GEP while the classical potential is bounded from above. To show that
$E_{0}$ is in fact bounded from below, we note that for $B\neq0$, one can get
the parametrization;
\begin{align}
B  &  =-\sqrt{\frac{M^{2}}{-4g}},\nonumber\\
M  &  =\sqrt[3]{6g}. \label{param}%
\end{align}
Accordingly, we get the relation $M=\frac{-3}{2B^{2}}$ and thus%

\[
E_{0}=\frac{1}{24B^{2}}\left(  8B^{6}g-9\right)  ,
\]
or in terms of a real parameter $b=\frac{B}{i}$;
\[
E_{0}=\frac{1}{24b^{2}}\left(  8gb^{6}+9\right)  ,
\]
which is bounded from below and positive (see Fig.\ref{eob}).

For $g$ positive, the condensate $B$ is pure imaginary and thus the
quasi-particle Hamiltonian in Eq.(\ref{quasip}) is $\mathcal{PT}-$symmetric
and up to first order one can get the metric operator in a simple fashion.
Indeed, the existence of a positive definite metric operator assures the \ the
reality of the spectrum. We will stress this point later in this work.\newline
Let us test the vacuum energy in comparison with its exact (numerical) and WKP
predictions from Ref.\cite{bendr} (table \ref{e0}). The calculations are
carried out at $g=1$ and the value in the table is $2E$ \ which is equivalent
to $E$ in Ref.\cite{bendr}.

\begin{table}[ptbh]
\begin{center}%
\begin{tabular}
[c]{|l|l|l|}\hline
Exact & our prediction (first order) & WKB\\\hline
1.4771 & $1.\,\allowbreak362\,8$ & 1.3765\\\hline
\end{tabular}
\end{center}
\caption{The first order $2E_{0}$ calculated at $g=1$ and compared to the
exact and WKP results from Ref.\cite{bendr}.}%
\label{e0}%
\end{table}One can realize that the effective field \ calculations though
simple are reasonable, taking into account that this is the first order
calculations and one can refine it by taking Feynman diagrams from higher
orders into account. Moreover, the effective potential is the generating
functional from which one can predict all the 1PI amplitudes. For instance at
$g=\frac{1}{2}$, we get the one point function $\langle0|\phi|0\rangle
=B=-1.\,\allowbreak019\,8i$ \ compared to the exact value $-0.97347i$ in Ref.
\cite{bendcal} which again assures the reliability of the effective field
calculation for the $\mathcal{PT}$-symmetric $\left(  -\frac{g}{4}\phi
^{4}\right)  $ theory.

The success of the effective field theory to predict the ground state energy
and the one point function of the $\left(  -\phi^{4}\right)  $ potential may
be thought as an accidental result and more tests are needed to support the
validity of the method as a whole. To do that, one may calculate the whole
spectrum using the quasi-particle Hamiltonian in Eq.(\ref{quasip}). This can
be easily obtained as
\begin{align*}
E_{n}  &  =\langle n|H|n\rangle=\left(  n+\frac{1}{2}\right)  M-\frac{1}%
{2}g\left(  3\frac{\left(  4\left(  n+\frac{1}{2}\right)  ^{2}+1\right)
}{8M^{2}}\right) \\
&  +\left(  -\frac{1}{2}M^{2}-3gB^{2}\right)  \frac{\left(  n+\frac{1}%
{2}\right)  }{M}+\left(  -\frac{1}{2}B^{4}g\right)  .
\end{align*}

\bigskip For $g=1,$ the values of the first four levels of $2E_{n}$ are listed
in table \ref{en}.

\begin{table}[h]
\begin{center}%
\begin{tabular}
[c]{|l|l|l|l|}\hline
n \ \  & $2E_{n}$ (Exact) & $E_{n}$ (our prediction) & $E_{n}$ (WKB)\\\hline
0 & $1.4771$ & $1.\,\allowbreak362\,8$ & $1.3765$\\\hline
1 & $6.0033$ & $6.\,\allowbreak333\,5\allowbreak$ & $5.9558$\\\hline
2 & $11.8023$ & $12.\,\allowbreak584$ & $11.7689$\\\hline
3 & $18.4590$ & $19.\,\allowbreak739$ & $18.4321$\\\hline
\end{tabular}
\end{center}
\caption{$2E_{n}$ at $g=1$ compared to the exact values and WKP predictions
from Ref.\cite{bendr}.}%
\label{en}%
\end{table}The above calculations are carried out in the non-Hermitian
representation. While the energy levels are the same in any representation,
the 1PI amplitudes do differ from representation to another. In fact, the
physical 1PI amplitudes have to be obtained by endowing the Hilbert space with
the inner product $\langle0|\hat{O}|0\rangle_{\eta_{+}}$, where $\eta_{+}$ is
the positive definite metric operator and $\hat{O}$ is the operator
representing the amplitude \cite{spect,spect1}. To obtain $\eta_{+}$, let us
rewrite the Hamiltonian in the form
\[
H=H_{0}-\frac{g}{2}B^{4}+\epsilon H_{I},
\]
where $\epsilon$ is used for book keeping purposes which will be sent to $1$
at the end of the calculations. Note that, the presence of the $-\psi^{4}$ in
the effective Hamiltonian in Eq.(\ref{quasip}) may lead to the conclusion that
this term is non-Hermitian. However, quantum field calculations uses a free
field representation or in a quantum mechanical language we are working in a
space of harmonic oscillator basis which are square integrable on the real
axis. Accordingly, the effective field theory in Eq.(\ref{quasip}) is a real
line problem and thus the operator $-\psi^{4}$ is Hermitian. Moreover, the
Hamiltonian in Eq.(\ref{quasip}) is in the quantum mechanical form;%
\[
\frac{p^{2}}{2}+\rho x+\alpha x^{2}+\beta x^{3}+\gamma x^{4},
\]
where the case of $\gamma<0$ has been investigated in a pure real line study
\cite{quart}. In this study, it has been shown that the Hamiltonian form is a
quasi-exactly solvable potential because the Hamiltonian can be written in a
Lie algebraic form \cite{quasi1, quasi2};%

\begin{equation}
H=\sum\limits_{a,b}C_{a,b}J^{a}J^{b}+\sum_{a}C_{a}J^{a},
\end{equation}
where $J^{i}$ is a set of first order differential operators which generate a
finite-dimensional Lie algebra. Accordingly, the first few levels can be
obtained exactly. Moreover, the study in Ref. \cite{quart}, showed that the
ground state \ function is square integrable provided that $\beta$ is pure
imaginary as in our case. In other words, rather than the original theory in
Eq.(\ref{x4}) which is a non-real line problem, the study supports our
prediction that the Hamiltonian in Eq.(\ref{quasip}) is a real line problem.
In fact, one can show that
\[
J=-\frac{1}{2}\frac{i-2\sqrt{g}B^{3}}{i},
\]
and since $J$ (called the spin, $J^{a},J^{b}$ are characterized by $J=$
$0,\frac{1}{2},1,.....J$) should be positive real then $B$ should be negative
and imaginary. This explains our choice of the negative root for $B$ in
Eq.(\ref{param}). In fact, a detailed analysis of the spectrum and wave
functions of the theory using the quasi-exactly solvable technique will follow
in another paper (in progress).

Now, since $\eta_{+}H\eta_{+}^{-1}=$ $H^{\dagger}$ where $\eta_{+}=\exp\left(
-Q\right)  $ and $Q=Q_{0}+\epsilon Q_{1}+\epsilon^{2}Q_{2}++\epsilon^{3}%
Q_{3}+......$ we can get%

\begin{align*}
H^{\dagger} &  =\exp(-Q)H\exp(Q)=H+[-Q,H]+\frac{[-Q,[-Q,H]]}{2!}\\
&  +\frac{[-Q,[-Q,[-Q,H]]]}{3!}+.....
\end{align*}
Up to first order in $\epsilon$ we get $\left(  H_{I}^{\dagger}-H_{I}\right)
=\frac{1}{2}[-Q_{1},H_{0}]\Rightarrow$ $[Q_{1},H_{0}]=2i\operatorname{Im}%
H_{I}$, where $\operatorname{Im}H_{I}$ is the imaginary part of $H_{I}$. This
leads to the first order $Q$ operator for the $\mathcal{PT}$-symmetric
quasi-particle Hamiltonian in Eq.(\ref{quasip}) of the form;\bigskip%

\begin{equation}
Q_{1}=-\frac{4\left(  \frac{ig}{2}4B\right)  }{3M^{4}}\Pi^{3}-\frac{2\left(
\frac{ig}{2}4B\right)  }{M^{2}}\frac{1}{3}\left(  \Pi\psi^{2}+\psi\Pi\psi
+\psi^{2}\Pi\right)  -\frac{2\left(  2giB^{3}\right)  }{M^{2}}\Pi.
\end{equation}
In fact, one can go beyond this order in a systematic way but it is out of the
scope of this work. What we need to clarify is that the traditional quantum
field effective theory can be applied to the $PT$-symmetric theories without
the need to the choice of a complex contour and try to investigate the theory
on this contour. In fact, this is \ an interesting result especially for
people who keep an eye on the Higgs mechanism for which the traditional
effective field theory is popular while the method of choosing a complex
contour is not willing to work in higher dimensions (quantum field theory).

\begin{figure}[ptb]
\begin{center}
\includegraphics [width=8cm,height=8cm]{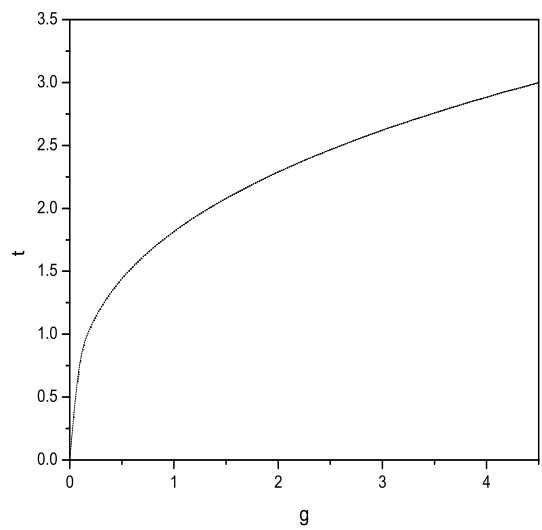}
\end{center}
\caption{The effective mass versus the coupling $g$ for the $\mathcal{PT}%
$-symmetric $-x^{4}$ potential.}%
\label{mass}%
\end{figure}

\begin{figure}[ptb]
\begin{center}
\includegraphics [width=8cm,height=8cm]{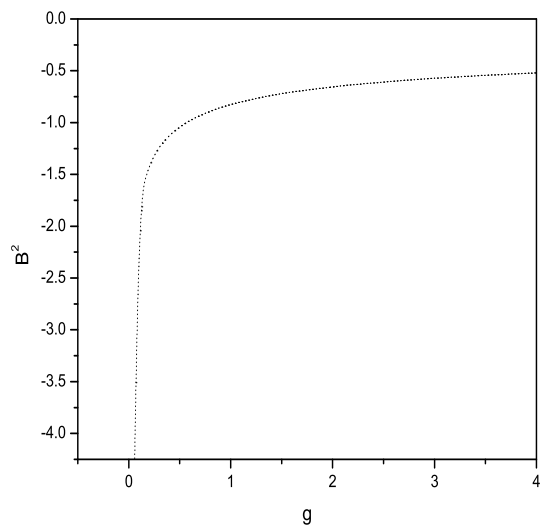}
\end{center}
\caption{The one point function $\left\langle 0| x|0\right\rangle $ squared
versus the coupling $g$ for the $\mathcal{PT}$-symmetric $-x^{4}$ potential.}%
\label{vaccond}%
\end{figure}

\begin{figure}[ptb]
\begin{center}
\includegraphics [width=8cm,height=8cm]{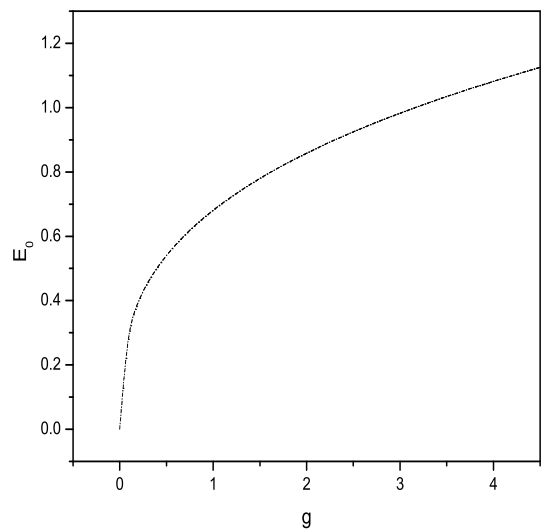}
\end{center}
\caption{The ground state energy versus the coupling $g$ for the
$\mathcal{PT}$-symmetric $-x^{4}$ potential.}%
\label{energy}%
\end{figure}

To test the range of the coupling for which the method is applicable to the case
under investigations, we plotted the effective field mass $M$, the vacuum
condensate squared $B^{2}$ and the ground state energy in Figs.\ref{mass}
,~\ref{vaccond} and \ref{energy}, respectively. In fact, the effective coupling
of the theory is $\frac{g}{M^{3}}=\frac{1}{6}$ \ \cite{efbook}. Accordingly,
the method is perturbative for the whole range of the coupling constant. In Figs.\ref{mass},
We can realize that the effective field mass parameter goes to zero as
$g\rightarrow0^{+}$ as expected. For the vacuum condensate, Fig. \ref{vaccond}, we plotted $B^{2}$
for positive values of the coupling $g$. In this figure,  $B^2$ is negative and thus the effective field theory is non-Hermitian and $\mathcal{PT}$-symmetric for
the $-\phi^{4}$ theory. Moreover, the condensate squared blows down as
$g\rightarrow0^{+}$, reflecting the full indeterminacy of position for the
free particle. To make this point clear, we consider the Hamiltonian form in
Eq.(\ref{quasip}) as $g\rightarrow0^{+}$. At this limit, $M$ $\rightarrow0$,
$2gB=g\sqrt{\frac{M^{2}}{-g}}=\sqrt{-gM^{2}}\rightarrow0$ and $gB^{3}%
=-\frac{3}{4}g^{\frac{2}{3}}\sqrt{-\frac{1}{\sqrt[3]{g}}}\rightarrow0$.
Accordingly, the Hamiltonian $H$ goes to $\frac{1}{2}\Pi^{2}$ which describes  a free particle and the corresponding wave solution is a plane
wave. Hence, the uncertainty in the position $x$ is $\Delta x=\sqrt
{\langle\left(  x-\langle x\rangle\right)  ^{2}\rangle}=\sqrt{\frac{1}{2M}%
}=\sqrt{\frac{B^{2}}{-\frac{1}{2}2^{\frac{2}{3}}3^{\frac{2}{3}}}}$, which is
known to be infinity for a plane wave and thus explains the big value of the
condensation as $g\rightarrow0^{+}$. Moreover, $\Delta p=\sqrt{\langle\left(
p-\langle p\rangle\right)  ^{2}\rangle}=\sqrt{\frac{M}{2}}=\sqrt{-\frac{1}%
{8}^{\frac{2}{3}}\frac{3^{\frac{2}{3}}}{B^{2}}}\rightarrow0$ as $g\rightarrow
0^{+}$ with the relation $\Delta x\Delta p=\frac{\hbar}{2}$ (\ $\hbar=1$) is
satisfied. Accordingly, in passing the tests of the axioms of quantum mechanics, our
calculations are proved to be valid.

In conclusion, we used an effective field theory formulation to calculate the
energy spectrum for the massless $-x^{4}$. The calculations showed that the
effective potential, up to first order in the coupling, is equivalent to
GEP. Moreover, the effective potential is bounded from below (Fig.\ref{eob}) and thus explains the stability of the energy spectrum which is against the naive classical analysis of concluding that the theory
is unstable as the classical potential is bounded from above.

From the vacuum energy (effective potential) one is able to predict all the
$1PI$ amplitudes. We compared our results with known exact numerical
calculations and though we have used only first order calculations of the
method, our results are in good agreement with the numerical calculations.
Moreover, we argued that the effective theory is perturbative for the whole
range of the coupling constant and the perturbation series is expected to
converge rapidly (the effective coupling $g_{eff}=\frac{1}{6}$).

The effective field theory introduced in Eq.(\ref{quasip}) is a general form
in the sense that it can be applied for any space time dimensions. What makes
this form interesting is that one can use it to calculate the  positive definite metric operator for the $\mathcal{PT}$-symmetric $-\phi^{4}$ \cite{abocop} which is a forward step toward a concrete formulation of a
$\mathcal{PT}$-symmetric Higgs mechanism. 

At the quantum mechanical level, the effective field theory introduced in
Eq.(\ref{quasip}) is a quasi-exactly solvable theory and thus one can obtain a
number of energy levels and wave functions exactly with the parameters $M$ and
$B$ are obtained from the effective potential. Such kind of calculations are
interesting and will follow in another paper.
\begin{acknowledgments}
We would like to thank Dr.~ B.~A. Shehadeh
 for his help in revising the manuscript.
\end{acknowledgments}

\newpage



\end{document}